\documentclass[letter]{aa} 
\usepackage{graphicx}
\usepackage{epsfig}
\usepackage{psfig}
\usepackage{txfonts}
\usepackage{natbib}
\bibpunct{(}{)}{;}{a}{}{,} 

\def\asec{\ifmmode ^{\prime\prime}\else$^{\prime\prime}$\fi}

\def\msun{\hbox{M$_{\odot}$}}

\def\degs{\ifmmode ^{\circ}\else$^{\circ}$\fi}
\def\amin{\ifmmode ^{\prime}\else$^{\prime}$\fi}
\def\asec{\ifmmode ^{\prime\prime}\else$^{\prime\prime}$\fi}

\def\degs{\ifmmode ^{\circ}\else$^{\circ}$\fi}
\def\amin{\ifmmode ^{\prime}\else$^{\prime}$\fi}
\def\EE#1{\times 10^{#1}}

\def\cm{\mbox{\,cm}}

\def\cm3{\mbox{\,cm$^{-3}$}}
\def\kms{\mbox{\,km~s$^{-1}$}}

\def\ergs{\mbox{\,erg~s$^{-1}$}}
\def\ergshz{\mbox{~erg~s$^{-1}$~Hz$^{-1}$}}
\def\kms{\mbox{\,km s$^{-1}$}}

\def\lsim{\!\!\!\phantom{\le}\smash{\buildrel{}\over
 {\lower2.5dd\hbox{$\buildrel{\lower2dd\hbox{$\displaystyle<$}}\over
                                 \sim$}}}\,\,}
\def\gsim{\!\!\!\phantom{\ge}\smash{\buildrel{}\over
{\lower2.5dd\hbox{$\buildrel{\lower2dd\hbox{$\displaystyle>$}}\over
                               \sim$}}}\,\,}

\begin{document}
 \title{Serendipitous discovery of the long-sought AGN in Arp 299-A} 
\titlerunning{Serendipitous discovery of the long-sought AGN in Arp 299-A}

   \author{M.A. P\'erez-Torres \inst{1}
      \and 
      A. Alberdi \inst{1}
      \and
      C. Romero-Ca\~nizales \inst{1}
     \and
     M. Bondi \inst{2}
                   }
          
   \institute{Instituto de Astrof\'{\i}sica de Andaluc\'{\i}a - CSIC, PO Box 3004, 18008 Granada,  Spain \\
             \email{torres@iaa.es}
         \and
         INAF-Istituto di Radioastronomia, Via Gobetti 101, I-40129, Bologna, Italy
            }

\date{Received 23 July 2010; Accepted 12 August 2010
}

\abstract
{The dusty nuclear regions of luminous infra-red galaxies (LIRGs) are
  heated by either an intense burst of massive star formation, an
  active galactic nucleus (AGN), or a combination of both.
  Disentangling the contribution of each of those putative dust-heating agents is a challenging task, and direct imaging of the
  innermost few pc can only be accomplished at radio wavelengths, 
  using very high-angular  resolution observations.}
{We observed the nucleus A of the interacting starburst galaxy
  Arp 299, using  very long baseline interferometry (VLBI)
  radio observations at 1.7 and 5.0 GHz.  Our aim was to characterize
  the compact sources in the innermost few pc region of Arp
  299-A, as well as to detect recently exploded core-collapse
  supernovae.  }
{We used the European VLBI Network (EVN) to image the 1.7 and 5.0 GHz compact radio emission of the
  parsec-scale structure in the nucleus  of Arp 299-A with milliarcsecond
  resolution.}
{Our EVN observations show that one of the compact VLBI sources, A1,
  previously detected at 5.0 GHz, has a flat spectrum between 1.7 and
  5.0 GHz and is the brightest source at both frequencies.  Our 1.7
  GHz EVN image shows also diffuse, low-surface brightness emission
  extending westwards from A1 and displays a prominent core-jet
  structure.
}
{The morphology, radio luminosity, spectral index and ratio of
  radio-to-X-ray emission of the A1-A5 region is consistent with a low-luminosity AGN (LLAGN), and rules out the possibility that it
  is a chain of young radio supernovae (RSNe) and supernova remnants
  (SNRs). We therefore conclude that A1-A5 is the long-sought AGN in
  Arp 299-A. This finding may suggest that both starburst and AGN are 
frequently associated phenomena in mergers.
}


   \maketitle
%

\section{Introduction}

Arp 299 consists of two interacting galaxies (IC 694 and NGC 3690) that are in an early merger stage \citep{keel95}. At a luminosity
distance of 44.8 Mpc \citep{fixsen96} for $H_0 =
73$~km~s$^{-1}$~Mpc$^{-1}$ (1'' corresponds to 217 pc) , Arp~299 has
an infrared luminosity, $L_{\rm IR} \approx 6.7\EE{11} L_{\odot}$
\citep{sanders03}.  The nuclear regions of Arp 299 are heavily
dust-enshrouded, which prevents their study at optical
wavelengths. Fortunately, radio observations are not affected by dust,
and therefore stand as the most promising way to study the innermost,
buried regions of Arp 299.

Radio interferometric VLA observations of Arp 299 at about one
arcsecond resolution (e.g. \citealt{neff04}) show that its radio
emission is mainly due to ionized gas. The large distance to Arp 299
requires the use of very long-baseline interferometry (VLBI) technique
to resolve the radio emission from the innermost regions of this
interacting starburst galaxy.  In fact, VLBI observations of the two
primary nuclei of Arp~299 have resulted in the discovery of a large
population of very compact sources in the nuclear regions of Arp~299-A
(=IC694) and Arp~299-B1 (=NGC 3690) \citep{pereztorres09a,ulvestad09}.
The high brightness temperatures observed for this population of
compact sources were indicative of a nonthermal origin for the
observed radio emission, implying that most of those sources were
young RSNe and SNRs.  On the other hand, X-ray imaging and
spectroscopy \citep{dellaceca02,zezas03,ballo04}, as well as the
presence of nuclear H$_2$O masers \citep{henkel05,tarchi07}, indicate
that the nuclear regions of Arp 299-A and Arp 299-B1 are likely to
contain an AGN, but its location has so far remained an open question.
On the other hand, \citet{almudena09} have noticed that the MIR properties of
Arp 299-A were consistent with obscured star formation, although they
could not exclude the existence of an LLAGN.  In this Letter, we focus
on studying the inner few parsecs of the nucleus in Arp 299-A. A
detailed discussion of the radio emission and spectral properties of
all the compact components in the central hundred parsec region will
be presented in another paper.

\section{EVN observations of Arp 299-A}

We used the European VLBI Network (EVN) to image the central region
of Arp~299-A at 1.7~GHz on 7-8 June 2009, and at
5.0 GHz on 12-13 June 2009, with the pointing position
centered on the Arp~299-A nucleus.  We used a sustained data recording
rate of 1024 Mbit s$^{-1}$ in two-bit sampling. Each frequency band
was split into eight intermediate frequencies (IFs) of 16 MHz
bandwidths each, 
for a total synthesized bandwidth of 128~MHz.  
Each IF was in turn split into 32 channels of 0.5~MHz bandwidth each.  
Our array included the following antennas at both frequencies:
Effelsberg, Cambridge, Medicina, Jodrell Bank (Mk 2), Onsala, Torun,
Urumqi, Shanghai, and Westerbork. In addition, the Yebes antenna in
Spain was also part of the array at 5.0 GHz.
Each of the two VLBI observing epochs consisted of 6.0 hr
phase-referenced experiments. Our target source, the nucleus of
Arp~299-A, was phase-referenced to the calibrator J1128+5925 with a
duty cycle of 5 minutes.  The strong source 4C~39.25 was used as
fringe finder and bandpass calibrator.  The total effective time on
source for Arp~299-A was  $\sim$3.6 hr.

The data were correlated at the EVN MkIV data processor at JIVE using
an averaging time of 2~s.
We performed standard a-priori gain calibration using the measured
gains and system temperatures of each antenna.  This calibration, as
well as the data inspection and flagging, were done within the NRAO
Astronomical Image Processing System (AIPS).  We also corrected for
ionosphere effects (of particular relevance at 1.7 GHz) and
source-structure effects of the phase-reference source at both
frequencies, following the same procedures as described in
\citet{pereztorres09a}. No self-calibration was performed on the data, since the peaks of emission
were too faint to qualify for these procedures.  We also used AIPS to
image Arp 299-A, that is, the images in Figure \ref{fig,arp299a} were
obtained by applying natural weighting to the data, which resulted in
FWHM, synthesized interferometric beams of (11.9 mas $\times$ 4.4 mas
at PA=-76$^\circ$) and (5.0 mas $\times$ 4.1 mas at PA=7$^\circ$) at 1.7 and
5.0 GHz, respectively.  The attained $1\sigma$ off-source r.m.s noise
was of $\sim 25\,\mu$Jy at 1.7 GHz and $\sim 22\,\mu$Jy at 5.0\,GHz.

\begin{figure}
\centering  
\includegraphics[width=90mm,angle=0]{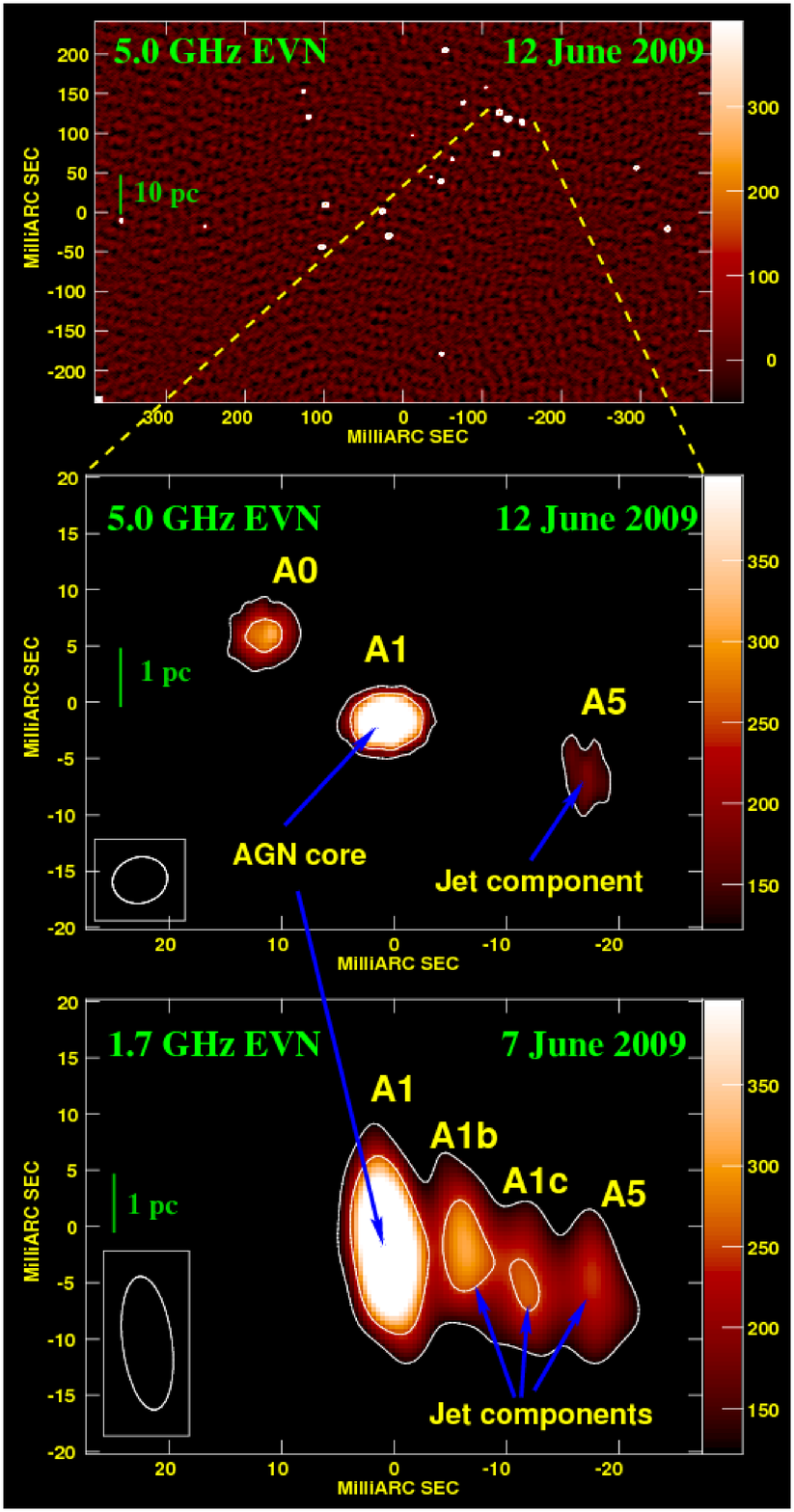} \\
\caption{ {\it Top:} 5.0 GHz full EVN image of the central 150 parsec
  region of the luminous infrared galaxy Arp 299-A (=IC 694), 
displaying a large number of bright, compact, nonthermal
  emitting sources, mostly identified with young RSNe and
  SNRs. The image center is at RA 11:28:33.63686 and
  DEC 58:33:46.5806.  {\it Middle and bottom:} Blow-ups of the inner
  8 parsec of the nuclear region of Arp 299-A, as imaged with the full
  EVN at 1.7 and 5.0 GHz.  The image center is at RA 11:28:33.61984
  and DEC 58:33:46.7006 in both panels. The morphology, spectral
  index and luminosity of the A1-A5 region are very suggestive of a
  core-jet structure.  The color scale goes from -50 $\mu$Jy/b up to 400
  $\mu$Jy/b in the top panel and from 125 $\mu$Jy/b to 400 $\mu$Jy/b in
  the middle and bottom panels. Contours are drawn at 5 and 10 times
  the off-source r.m.s. noise.}
 \label{fig,arp299a}
\end{figure}

\section{Results and discussion}

Our results are summarized in Figure \ref{fig,arp299a} and Table
\ref{tab,evn}.  Figure \ref{fig,arp299a}  shows the field
of the central (150$\times$80) pc of Arp 299-A (top), as imaged with
the EVN at 5.0 GHz on 12-13 June 2009.  Our 5.0 GHz EVN image 
is very similar to the images previously published in
\cite{pereztorres09a}, and confirms that Arp 299-A hosts a large
number of bright, compact radio sources. As already noted by
\cite{pereztorres09a}, the high brightness temperatures displayed by
those compact components indicate that they must be nonthermal
emitting sources, most likely young radio SNe and SNRs, and a detailed
discussion of the radio emission and spectral properties of all these
compact sources will be presented elsewhere.  The middle and bottom
panels of the figure are blow-ups of the inner 8 parsec nuclear region
of Arp 299-A, as imaged with the full EVN at 1.7 and 5.0 GHz, and the
positions of its compact components, as well as their flux densities, 
are summarized in Table \ref{tab,evn}.

The main results from our observations concern the nature of component A0 and,
especially, of the complex formed by A1-A5 (lower panel in Figure \ref{fig,arp299a} and
Table \ref{tab,evn}), which we discuss in the next sections.


\begin{table*}[htbp]
\centering                          
\caption{The parsec-scale structure of Arp 299-A as imaged with the EVN}             
\label{tab,evn}      
\renewcommand{\footnoterule}{}
\begin{tabular}{llcclrrl}
\hline \hline
Source & Source & $\Delta\,\alpha^{\mathrm{a}}$ & $\Delta\,\delta^{\mathrm{a}}$ &
Angular size$^{\mathrm{b}}$ & S$_{1.7}$ ($\mu$Jy) $^{\mathrm{c}}$ 
               & S$_{5.0}$ ($\mu$Jy)$^{\mathrm{c}}$ &
               Spectral index  \\
Name  & Type & (J2000.0)  & (J2000.0)  & (milliarcsec) &  & & $\alpha_{1.7}^{5.0}$ \\
 \hline 

A0  &      SN & 33.6213 & 46.707 & $\leq 4.1\times4.0$ &    $\leq$75 &307$\pm$27 &  $\geq$1.28  \\
A1   &   AGN (core)    & 33.6199 & 46.699 & $\leq 6.1\times 2.7$  & 822$\pm$48 & 716$\pm$42 & -0.13$\pm$0.11\\
A1b &  AGN (jet)     & 33.6190 & 46.699 & $\leq 10.1\times 5.1$  &314$\pm$30 & $\leq$ 66 &  $\leq$-1.42  \\
A1c &  AGN (jet)     & 33.6183 & 46.695 &  $\leq 14.3 \times 6.2$
& 267$\pm$28 &  $\leq$ 66 & $\leq$-1.27 \\
A5  &   AGN (jet)        & 33.6176 & 46.694 & $\leq 6.9 \times 5.0$     &246$\pm$28  & 183$\pm$24  &  -0.27$\pm$0.22  \\

\hline
\end{tabular}   
\begin{list}{}{}
\item[$^{\mathrm{a}}$] Coordinates are given with respect
  to $\alpha$= 11:28:00.0000 and $\delta$ =
  58:33:00.000, and the typical uncertainty is less than 1 mas.
\item[$^{\mathrm{b}}$] Upper limits to the deconvolved angular size (Major axis $\times$
  Minor axis, in mas) as obtained using  task IMFIT within AIPS, using
  our 5.0 GHz for all components but A1b and A1c, which were only
  detected at 1.7 GHz.
\item[$^{\mathrm{c}}$] The uncertainty in the reported flux density
  (and derived spectral index) for the detected compact components
  corresponds to 1$\sigma$, where $\sigma$ was determined by adding in
  quadrature the off-source r.m.s.  in each image and a 5\% of the
  local maxima, to conservatively account for possible inaccuracies in
  the EVN calibration. Upper limits for undetected sources are three
  times the off-source r.m.s.
\end{list}
\end{table*}

\subsection{The young radio supernova A0}

Component A0 was identified with an RSN by
\citet{neff04}. A0 showed a strongly inverted spectral index between
2.3 and 8.4 GHz, indicating that it was a very young SN.
We already detected it at 5.0 GHz \citep{pereztorres09a}, more than
five years after its discovery at 8.4 GHz, which implied that A0 was a
longlasting, slowly evolving, nonstandard RSN
\citep{pereztorres09a}.  Our current 5.0 GHz EVN observations show
that A0 continues to be very bright, with $L_{5.0} = 7.4
\EE{26}$\ergshz, confirming that it is a longlasting, slowly evolving
Type II supernova in a rich, dense surrounding medium. Not
detecting A0 at 1.7 GHz also confirms the suggestion by
\cite{pereztorres09a} that its radio emission is being suppressed at
low frequencies by a nearby absorber, most likely a foreground H~II
region in the vicinity of A0. We can characterize the main parameter
of this H~II region, its emission measure $EM$, by assuming an
(intrinsic) spectral index typical of Type II SNe in their optically
thin phase [$\alpha = -0.75$; $S_{\rm intr, 1.7} = S_{\rm 5.0} (1.7/5.0)^\alpha$]; 
and purely foreground, free-free absorption  
($S_{\rm obs, 1.7} = S_{\rm intr, 1.7} \, e^{-\tau_{\rm ff, 1.7}}$).  
We then obtain $\tau_{\rm ff, 1.7} \geq$ 2.2.  The $EM$ at GHz frequencies is proportional to
the free-free optical depth and is given by $EM = 3.1\EE{6} \, \tau_{{\rm ff}, \nu}\,
T^{1.35}_{e,4} \, \nu^{2.1}$ cm$^{-6}$ pc  \citep{mezger67}, where
$T_{e,4}$ is in units of $10^4$ K, and $\nu$ in GHz.  Assuming 
the free-free absorber is a purely hydrogen plasma with $T_{e,4} = 1$,
we obtain $EM \geq 2.0
\EE{7}$ cm$^{-6}$ pc for the H~II close to A0. This value of $EM$ is very similar to the one obtained
for the foreground H~II region close to the well studied
SN~2000ft \citep{alberdi06,pereztorres09b}, in the dust-enshrouded
galaxy NGC 7469.

\subsection{The A1-A5 complex}

We detected components A1 and A5, which were already reported at 5.0
GHz by \citet{pereztorres09a}. A1, the brightest compact source
detected during our previous 5.0 GHz eEVN observations, continues to
be the brightest one at 5.0 GHz ($L_{5.0} = 1.8 \EE{27}$\ergshz), and
is also detected at 1.7 GHz ($L_{1.7} = 2.0 \EE{27}$\ergshz).
Component A5 is also detected at both frequencies.
\citet{pereztorres09a} suggested that A1 could be an SNR, while A5 it was
suggested a possible SN, based on its marginal variability at
5.0 GHz (although we could not exclude that A5 was an SNR).  However,
our new, contemporaneous observations at 1.7 and 5.0 GHz shed new
light on their nature, and suggest a rather different scenario.

In fact, A1 and A5 are part of a ``global structure'', as clearly
shown by our 1.7 GHz EVN observations (see lower panel in Figure
\ref{fig,arp299a}), where a bridge of radio emission connects them,
with two new regions of enhanced emission in between (components A1b
and A1c) .  The extension of the A1-A5 complex is $\simeq$
31$\times$21 mas, corresponding to a linear size of $\simeq$
6.7$\times$4.6 pc.  Within the 1.7 GHz structure, we detect two new
components, A1b and A1c, which are not detected at 5.0 GHz more than three
times the off-source r.m.s. Nevertheless, detecting them clearly at 1.7 GHz
($\geq 9\sigma$) implies that they are real, steep-spectrum
components, whose nature we discuss in the next two sections.

\subsubsection{A chain of nested SNe and SNRs in an SSC...?}

The spectrum of A1 between 1.7 and 5.0 GHz is very flat ($\alpha =
-0.13 \pm 0.11$). This spectral index would in principle be compatible
with an H~II region but, as we already showed in
\citet{pereztorres09a}, all compact components in Arp~299-A show
brightness temperatures that imply a nonthermal origin for their
radio emission. Thus, the flat spectrum shown by A1 could be explained
by radio emission from an SNR or, alternatively, from an AGN.  How can
we distinguish these two scenarios? To answer this question we
must consider A1 in the context of the complex it is part of, together
with components A1b, A1c, and A5.  As with A1, the brightness
temperatures of the other components rule out a thermal origin for
their radio emission. In principle, the spectral indices of those components (see
Table \ref{tab,evn}) are compatible with A1b and A1c being young
supernovae in their optically thin phase, and A5 being another SNR,
all of them part of a super star cluster (SSC).

The existence of SSCs in Arp~299-A has been demonstrated by their
detection using 2.2-$\mu$m adaptive optics imaging
\citep{lai99}, and  further evidence comes from {\it
  Hubble Space Telescope\/} ({\it HST}\/) FOC and NICMOS images, which
reveal a population of young  (4$-$15 Myr) stellar clusters in the central regions
of Arp 299-A \citep{almudena00}. 
High-resolution near-infrared spectra of the nuclear starburst of M~82 \citep{mccrady03}
have shown that its SSCs have typical radii of 2$-$3 pc, and masses that range from 
$\sim 10^5$ \msun\, up to (1-2)$\EE{6}$ \msun.
The size of the A1-A5 complex is 6.7 $\times$ 4.6 pc, in agreement
with expectations for SSC sizes.  Therefore, a possible scenario is
one where a young SSC has resulted in two young, recently exploded SNe
(A1b and A1c), and two SNRs (A1 and A5). In this case, the steepness
of the spectral indices for sources A1b and A1c can only be explained
as from very recently exploded ($t \lsim 10-20$ yr) core-collapse
supernovae. Furthermore, to estimate the age of the SNR candidates A1 and
A5, we use upper limits to their (deconvolved) angular size (column 5
in Table \ref{tab,evn}).  A1 and A5 have a maximum angular radius of 6.1
and 6.9 mas, which correspond to a linear size of 1.3 and 1.5 pc,
respectively. Assuming an average expansion speed of 3000 \kms for
both SNRs, those sizes correspond to kinematic ages of $\simeq$430 yr
for A1 and $\simeq$490 yr for A5.

In summary, if the A1-A5 complex is an SSC, it has yielded four bright
radio supernovae in the past 500 yr. Now, this is very unlikely to
happen in an SSC of just 3.4 pc radius whose mass is unlikely to be
higher than (1-2)$\EE{6}$ \msun. Indeed, evolutionary models for the
radio emission in starbursts by \citet{perezolea95} show that the SN
rate in an instantaneous burst can be as high as 20$\EE{-10}$
yr$^{-1}$\msun$^{-1}$ up to an age of $\simeq$6 Myr, and about
10$\EE{-10}$yr$^{-1}$\msun$^{-1}$ at an age of 9-15 Myr. Thus, an SSC
of 10$^6$\msun would yield $\leq$1 SN every 500 yr at an age of 6 Myr,
and $\leq$1 SN every 1000 yr at an SSC age of 9-15 Myr. Therefore, if
A1-A5 were SNe and SNRs exploding in a SSC, we should have expected to
detect at most one SN/SNR in the past 500 yr, whereas we find
four.  The probability of detecting two (young) SNe in the past 20 yr,
and two SNRs with ages of $\sim$430 and $\sim$ 490 yr in the past 500
yr is less than $3\EE{-6}$.  All this evidence seems to rule out an SSC
scenario for the A1-A5 complex.

\subsubsection{...Or the long-sought AGN in Arp 299-A?}

We show in the next paragraph that a much more plausible scenario for the A1-A5
complex is that of an AGN, where A1--the brightest compact source at
both frequencies in the nuclear and circumnuclear regions of Arp
299-A-- is identified with the core of the AGN in Arp
299-A, while components A1b, A1c, and A5 are jet components (see Figure
\ref{fig,arp299a}). There are strong grounds for this scenario:

\noindent (i) A1-A5 displays a core-jet morphology, typical of AGNs.
(ii) The two-point spectral indices of the A1-A5 complex (see Table~\ref{tab,evn})
agree well with expectations from an AGN. In this scenario A1, which has the
flattest spectrum between 1.7 and 5.0 GHz, would be the core, and A1b,
A1c, and A5 would be jet components.
(iii) There is evidence of strong flux density variability of A1 at
2.3 and 8.4 GHz between 2002 and 2003 \citep{neff04, ulvestad09}.
Figure 5 of \citet{ulvestad09} shows little variation for A1 starting
from 2004, or even a slight decrease (increase) in flux density at 2.3
(8.4) GHz. On the other hand, there is no evidence for significant
variability at 5.0 GHz between April 2008 and June 2009 (see
\citealt{pereztorres09a} and Table~\ref{tab,evn}).  This behavior is
at odds with an SN origin for the radio emission of A1, and is more
easily reconciled within an AGN scenario.  In addition, by combining the data from
\citet{ulvestad09} with ours and assuming no variability between 2005
and 2009 at cm-wavelengths, we find that the spectrum of A1 must
have a turnover frequency between 1.7 and 5.0 GHz, as expected for the
core of an AGN whose radio emission is partially self absorbed.
(iv) The radio luminosity of the A1-A5 complex between 1.7 and 8.4 GHz
is $\simeq 1.9\EE{37}$\ergs, typical of LLAGNs.  (v) The ratio of (5
GHz) radio-to-X ray (soft) luminosity (\citealt{terashima03}) for the
A1-A5 complex in Arp 299-A is typical of LLAGNs.  From Table
\ref{tab,evn}, we obtain $\nu\, L_{\nu, 5.0} \approx 2.0 \EE{37}$
\ergs. The combined absorption corrected X-ray
luminosity for the X-ray sources discovered by \citet{zezas03} in Arp
299-A is $\sim 1.9 \EE{40}$\ergs.  This results in a value of $R_{\rm
  X} \simeq 1.1\EE{-3}$, which agrees closely with expectations from an
LLAGN (see Figure 4 of \citealt{terashima03}).

We therefore identify the A1-A5 complex with the AGN in Arp 299-A. In
this case, the cm-radio luminosity of the AGN in Arp 299-A is very
similar to the one displayed by the LLAGN in M~81, which is the closest
galaxy with an active nucleus. The core of M~81 also presents a faint,
one-sided jet, with a flat spectrum near the core and a steep spectrum
along the jet \citep{bietenholz04}.  
Furthermore, cm-VLBI observations of the merger system
NGC 6240 \citep{gallimore04} and
mm-Plateau-de-Bure interferometer observations of the prototypical
ULIRG Arp 220 \citep{downes07} have suggested the coexistence of both
AGN and starburst in their nuclear regions.
Thus, our discovery suggests that both starburst and AGN are frequently associated in merging systems.

\section{Summary}

We imaged with milliarcsecond resolution the inner eight parsecs of the
nuclear region in Arp 299-A, using contemporaneous EVN observations at
1.7 and 5.0 GHz.  At 5.0 GHz, we detect components A0, A1, and A5, as
previously detected by \citet{pereztorres09a}. At 1.7 GHz, A1 and A5
are also detected, but not A0. In addition, two new components  between A1 and
A5 (A1b and A1c) are detected at 1.7 GHz, forming a complex.

We find that the morphology, spectral index, radio luminosity, and
radio-to-X ray luminosity ratio of the A1-A5 complex are consistent
with that of an LLAGN, and rules out the possibility that it consists
of a chain of young RSNe and SNRs in a young SSC. We therefore
conclude that A1 is the long-sought AGN in Arp 299-A.
Since Arp 299-A had long been thought of as a pure starburst, our finding of a buried, low-luminosity AGN in its central region, coexisting with a recent burst of starformation, suggests that both a starburst and AGN are frequently associated phenomena in mergers. In this case, our result is likely to have an impact on  
evolutionary scenarios proposed for AGN and the triggering mechanism of activity in general.

Finally, we also note that component A0, previously identified as a
young RSN, is not seen at our low-frequency observations,
which implies there is a foreground absorbing H~II region.  It is
remarkable that this RSN exploded at the  mere distance
(projected) of two parsecs from the putative AGN in Arp 299-A, which
makes this supernova one of the closest to a central supermassive
black hole ever detected. This result may also be relevant to
accreting models in the central regions of galaxies, since it is not
easy to explain the existence of very massive, supernova progenitor
stars so close to an AGN. While seemingly contradictory, this could
explain the low-luminosity of the AGN we see in Arp 299-A. In
fact, since massive stars shed large amounts of mechanical energy into
their surrounding medium, thereby significantly increasing its
temperature, those massive stars would hinder the accretion of material to the
central black hole, which could in turn result in a less powerful AGN
than usual. 

\begin{acknowledgements}
We are grateful to  Luis Colina and Emilio Alfaro for useful discussions and to the anonymous referee for suggestions and comments that have improved the science of our paper.
 The EVN is a joint facility of European, Chinese, South African and
other radio astronomy institutes funded by their national research
councils.  MAPT, AA, and CRC acknowledge support by the Spanish
MICINN through grant AYA2009-13036-CO2-01. This research has been
also partially funded by the Autonomic Government of Andalusia under
 grants P08-TIC-4075 and TIC-126. 
 MAPT is also grateful for  financial support from the Spanish Research Council (CSIC) through
  grant 200950I139.
\end{acknowledgements}

\bibliographystyle{aa}
\bibliography{15462bib}

\Online

\end{document}